\documentclass[english,epj]{svjour}
\usepackage[T1]{fontenc}
\usepackage[latin9]{inputenc}
\usepackage{color}
\usepackage{amsmath}
\usepackage{amssymb}
\usepackage{graphicx}

\makeatletter
\usepackage{ulem}
\institute{Laboratoire PhLAM, CNRS UMR 8523, Bât. P5 - Université Lille1, 59655 Villeneuve d'Ascq cedex, FRANCE}
\abstract{The cloud of cold atoms produced by a Magneto-Optical Trap is known to exhibit instabilities. We examine in this paper in which limits it could be possible to realize an experimental trap similar to the configurations studied theoretically, i.e. mainly traps where one direction is privileged. We study the static behavior of an anisotropic trap, where anisotropy results essentially from the use of two different laser frequencies for the arms of the trap. Such a trap has very surprising behaviors, in particular the cloud disappears for some laser frequencies, while it exists for smaller and larger frequencies. A model is build to explain these behaviors. We show in particular that, to reproduce the experimental observations, the model has to take into account the cross saturation effects. Moreover, the couplings between the different directions cannot be neglected. 
}
\usepackage{cite}

\makeatother

\usepackage{babel}
\begin{document}

\title{The Dual Frequency Anisotropic Magneto-Optical Trap}

\author{Rudy Romain, Philippe Verkerk and Daniel Hennequin}

\maketitle

\section{Introduction}

The spectacular results obtained during the last decades in the domain
of the experimental quantum physics required all as a first step to
cool atoms with a Magneto-Optical Trap (MOT). The cold atoms produced
in such a MOT can then be put in lattices \cite{lattices}, used to
produce cold molecules \cite{molecules} or be further cooled down
to produce Bose-Einstein condensates \cite{BEC}. But the MOT itself
is also an interesting object. Many questions remain unanswered, and
the usual theoretical descriptions are hardly enough to describe the
stationary behavior of the cloud of cold atoms produced by a MOT.
But in many situations, this cloud exhibits spatio-temporal instabilities
\cite{wilkowski2000,labeyrie2006}, for which the development of new
models is necessary. And indeed, several models were recently proposed
\cite{wilkowski2000,labeyrie2006,hennequin2004,distefano2003,distefano2004,pohl2006,mendonca2008,romain2011}.
The search for a model describing correctly the dynamics of cold atoms
in a MOT is primarily motivated by the hope to understand the mechanisms
leading to such a complex behavior. But it regained recently interest
when it was shown that this system is formally very close to some
plasmas. In particular, it was demonstrated that the cloud of cold
atoms is described by a Vlasov-Fokker-Planck equation, and the analogies
with plasmas were extensively discussed \cite{romain2011}.

Unfortunately, none of these models lead to a satisfactory description
of the experimental observations. For most of them, the theoretical
predictions differ deeply from the experimental observations \cite{labeyrie2006,pohl2006,mendonca2008}.
Some models give a good qualitative description of the observed dynamics
\cite{wilkowski2000,hennequin2004}. But they modelize a 1D MOT, and
direct quantitative comparison with experiments is not possible, as
all experimental observations of the MOT dynamics concern 3D MOTs.
Thus, to go further, it is necessary either to generalize these models
in 3D, or to realize 1D experiments.

A 1D MOT in this context is more precisely a MOT where atoms are trapped
in 3D, but where instabilities occur only along one direction of space,
called in the following the unstable direction. Such a MOT is an anisotropic
MOT rather than a 1D MOT. It could be obtained if the unstable direction
of space is not coupled to the other directions. Unfortunately, the
effective coupling between the different arms of a MOT has been poorly
studied, while several mechanisms are known to possibly induced such
a coupling. For instance, the well known multiple scattering has never
been studied from this point of view. The cross saturation effects
are also a possible coupling mechanism, which is usually neglected.
It is important to evaluate precisely these couplings, as, if they
cannot be neglected, other solutions have to be used to force the
system to be stationary in two directions of space.

Various anisotropic traps have been studied in the past. The interesting
configurations for our purposes are those where the anisotropy is
introduced on a parameter controlling the instabilities of the MOT.
In \cite{distefano2003,distefano2004}, it is shown that at least
two control parameters allow to tune the trap from a stationary situation
to an unstable dynamics: the intensity of the laser trapping beams,
and the detuning between the laser trapping beam frequency and that
of the atomic transition used to cool the atoms\textcolor{black}{.
The main difference between these two control parameters concerns
the range on which occurs the transition from a stable behavior to
instabilities. The results in \cite{distefano2004} show that for
intensity, this range spreads over almost one order of magnitude,
whereas for the detuning, it is less than a factor 2.}

Traps with anisotropic laser beam intensities or magnetic field gradients
have already been studied \cite{stites2004,vengalattore2003}. One
of the main results is that for clouds with a large aspect ratio,
multiple scattering disappears \cite{vengalattore2003}. This puts
in evidence the coupling between the different directions of space
through the multiple scattering, and this means that the anisotropy
should be as small as possible to limit the effect of this coupling.
Thus in our case, introducing the anisotropy through the frequency
detuning should be a better solution, but to our knowledge, this type
of anisotropic trap has not yet been studied.

We present in this paper results about a dual frequency trap with
different frequency detunings along the different axes. Experimental
measurements show that such a trap has very unusual behaviors, in
particular the disappearance of the atomic cloud for some detuning
pairs, while cold atoms are obtained on both sides of these frequencies.
We show that the usual models, which neglect the cross saturation
effects, are not able to reproduce these experimental observations.
We build another model which takes into account the cross saturation
effects, and show that the behaviors predicted by this model are qualitatively
consistent with the experimental observations.

\section{Experimental results}

\subsection{Experimental setup}

We work with a Cesium-atom MOT in the usual $\sigma^{+}-\sigma^{-}$
configuration. Each of the three arms of the trap is formed by counter-propagating
beams resulting from the reflection of the three forward beams, obtained
from the same laser diode. Beams propagate following three perpendicular
directions, one of them being the axis of the coils producing the
magnetic field. In this so-called parallel direction, the forward
beam is characterized by the intensity $I_{\parallel}$ and the detuning
$\Delta_{\parallel}=\omega_{\parallel}-\omega_{0}$, where $\omega_{\parallel}$
is the beam frequency and $\omega_{0}$ the atomic frequency. In the
two other directions, the beams are characterized by the intensity
$I_{\perp}$ and the detuning $\Delta_{\perp}=\omega_{\perp}-\omega_{0}$.
When $I_{\parallel}=I_{\perp}$ and $\Delta_{\parallel}=\Delta_{\perp}$,
we have the most common MOT. However, this usual MOT is already not
isotropic, although it is often considered as so. Indeed, the magnetic
field is produced by a pair of coils in an anti-Helmholtz configuration.
Thus the magnetic field gradient along the parallel direction is twice
that in the perpendicular directions, leading to a fixed anisotropy
on the restoring force of the trap. We can characterize such a trap
as a balanced single-frequency trap. As discussed above, we will focus
here on the effects induced by a frequency anisotropy, i.e. a balanced
or unbalanced dual-frequency trap ($\Delta_{\parallel}\neq\Delta_{\perp}$). 

In the present study, we focus on the stationary cloud. This cloud
can be characterized by its number of atoms and the spatial distribution
of these atoms. To measure the number of atoms, we just need a photodiode
to record the total fluorescence emitted by the atoms in the cloud.
Fluorescence is proportional to the number $N$ of atoms in the cloud,
and thus it is a good indicator of $N$. However, the proportionality
coefficient depends on the laser intensities and detunings. For large
saturations, this coefficient may be considered as constant, but in
the other cases, it decreases as the detunings increase. In the experimental
results presented here, taking into account this correction factor
would lead to an increase of $N$ by a factor lower than 3 for large
detunings. As this correction induces no qualitative change in the
behaviors described here, we did not apply it on the curves shown
below.

The distribution of atoms is usually measured through the radius of
the cloud. However, we introduce now a new anisotropy in the trap,
which is expected to lead to an ellipsoidal cloud. Thus a measurement
of the cloud size in the two main directions appears necessary. This
measurement is performed by using a camera. Recorded pictures are
analyzed by a software which fits the atomic cloud on a 2D gaussian.
The result of the fit gives us the semi-axes $L_{\parallel}$ and
$L_{\perp}$ of the ellipsoid, but a convenient value to monitor is
the ellipticity $\varepsilon=L_{\perp}/L_{\parallel}$, as we expect
a correlation between $\varepsilon$ and the anisotropy.

\subsection{Atomic cloud behavior}

We have studied the evolution of the cloud as a function of the frequency
difference $\Delta_{\parallel}-\Delta_{\perp}$, for different detunings
and trap beam intensities, including cases where intensities following
the two directions are different. A typical experimental measurement
consists in recording the evolution of the number $N$ of atoms in
the cloud and the cloud sizes as a function of $\Delta_{\parallel}$,
the other parameters, in particular $\Delta_{\perp}$, $I_{\perp}$
and $I_{\parallel}$, being constant. The measurements are then repeated
for different values of $\Delta_{\perp}$, $I_{\perp}$ and $I_{\parallel}$.

\begin{figure}
\centering{}\includegraphics[width=8cm]{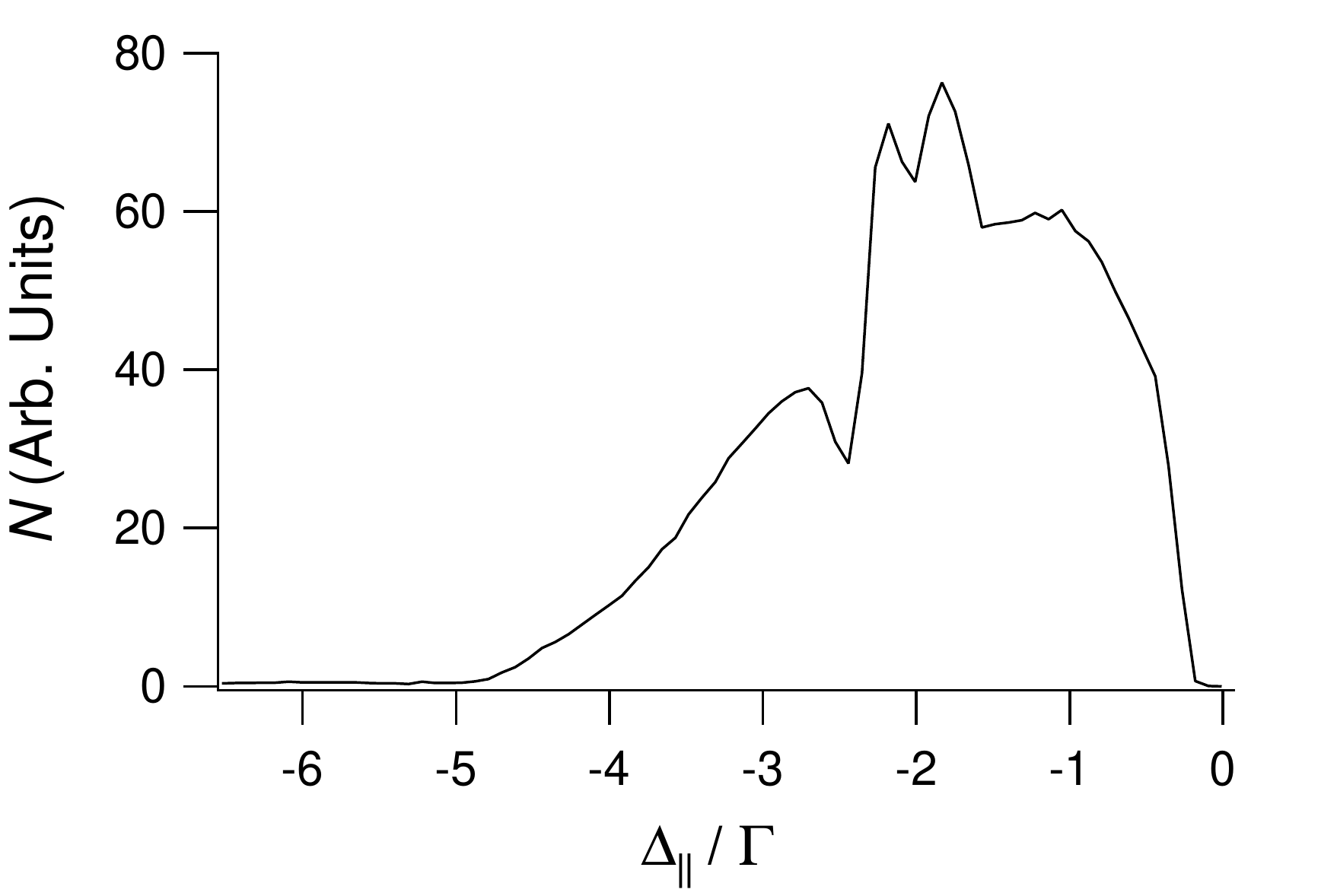}\caption{\label{fig:fullscale}Number $N$ of atoms in the cloud versus $\Delta_{\parallel}$
for $\Delta_{\perp}=-2\Gamma$ and $I_{\parallel}\simeq I_{\perp}\simeq2I_{S}$.}
\end{figure}

Fig. \ref{fig:fullscale} shows the typical evolution of the fluorescence
on the interval where it is measurable. In this example, $\Delta_{\perp}=-2\Gamma$,
where $\Gamma=2\pi\times5.234\,\mathrm{MHz}$ is the natural width
of the transition, $-6.5\Gamma<\Delta_{\parallel}<0$ and $I_{\parallel}\simeq I_{\perp}\simeq2I_{S}$
where $I_{S}=1.1\,\mathrm{mW/cm^{2}}$ is the saturation intensity.
At the degeneracy $\Delta_{\parallel}=\Delta_{\perp}$, we have a
standard balanced single-frequency MOT.

On each side of this degeneracy, at a typical frequency difference
$\Delta_{\parallel}-\Delta_{\perp}=\pm0.5\Gamma$, the number of atoms
exhibits a gap. We did not study in details the mechanisms at the
origin of this behavior, mainly because in the scope of the present
work, it is more interesting to introduce a large difference between
$\Delta_{\parallel}$ and $\Delta_{\perp}$, to obtain a large anisotropy.
However, it can be noticed that the frequencies where the gaps occur,
are of the same order of magnitude as the energy shifts between the
ground state Zeeman sublevels \cite{grison1991}. Thus it is probable
that this behavior originates in a Raman resonance between Zeeman
sub-levels of the fundamental atomic level.

\begin{figure}
\centering{}\includegraphics[width=8cm]{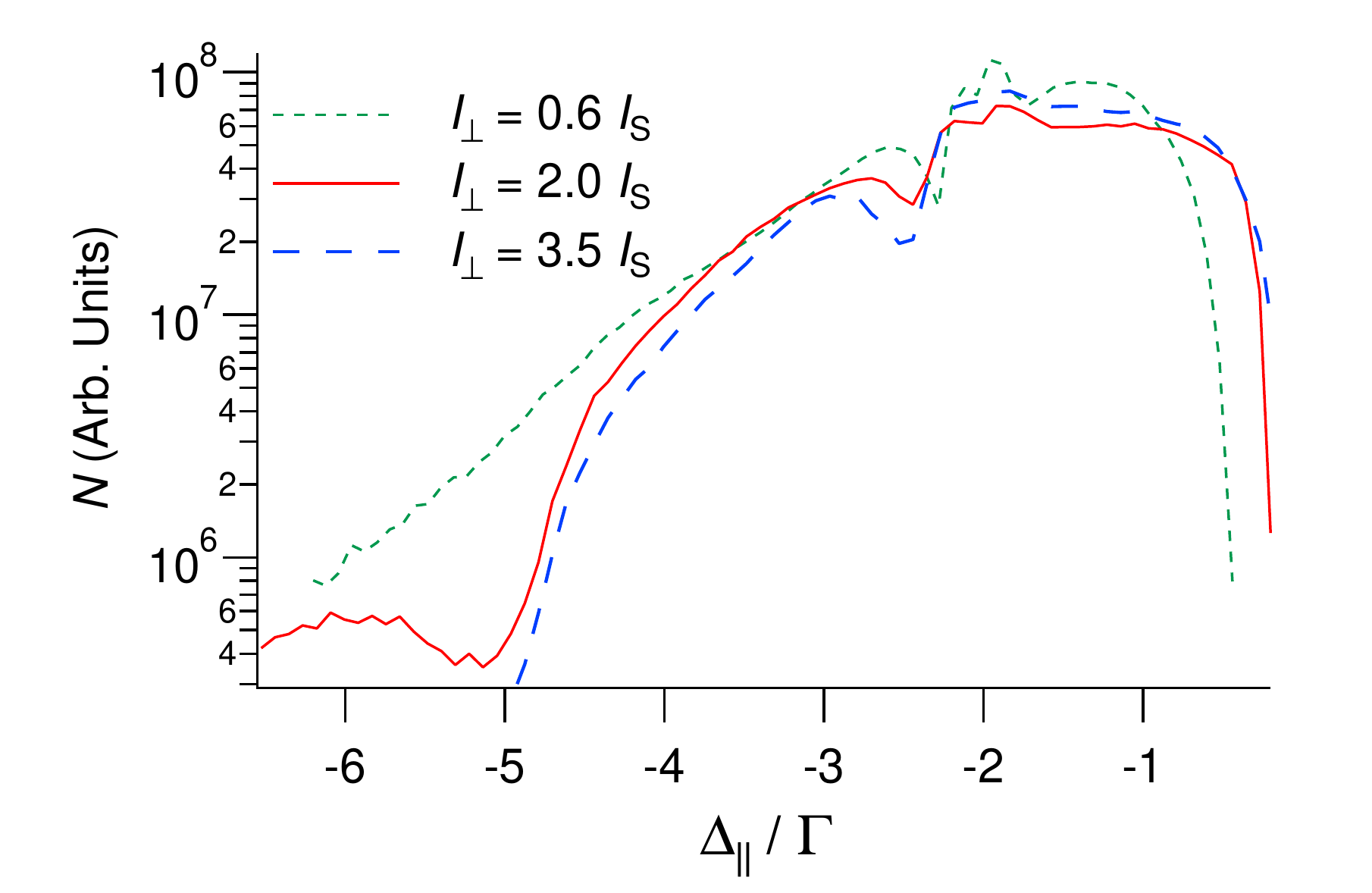}\caption{\label{fig:scenario2}Number $N$ of atoms in the cloud versus $\Delta_{\parallel}$
for $\Delta_{\perp}=-2\Gamma$, $I_{\parallel}\simeq2I_{S}$ and different
values of $I_{\perp}$.}
\end{figure}

\begin{figure}
\begin{centering}
\includegraphics[width=8cm]{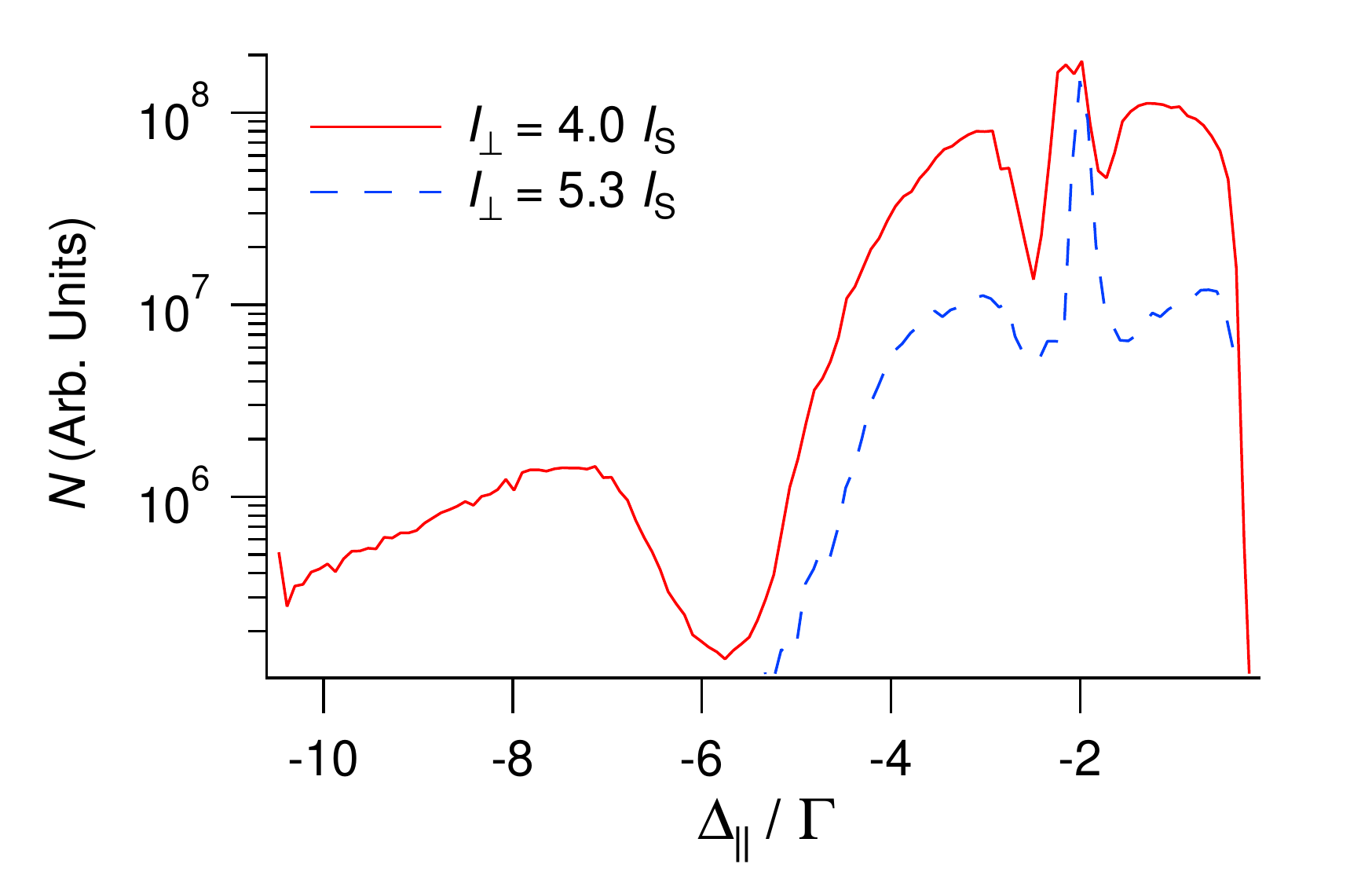}
\par\end{centering}

\caption{\label{fig:scenario10}Number $N$ of atoms in the cloud versus $\Delta_{\parallel}$
for $\Delta_{\perp}=-2\Gamma$, $I_{\parallel}\simeq9I_{S}$ and different
values of $I_{\perp}$.}

\end{figure}

Except for these gaps, the number of atoms in the MOT decreases monotonically
as $\left|\Delta_{\parallel}-\Delta_{\perp}\right|$ is increased.
On the blue side, it vanishes when $\Delta_{\parallel}$ reaches the
resonance, as expected when atoms are no more trapped in the parallel
direction. On the red side, we observe different behaviors, depending
on the laser intensities. More precisely, the observed behavior depends
mainly on $I_{\perp}$, and thus we find similar evolutions in traps
with $I_{\parallel}=I_{\perp}$ or $I_{\parallel}\neq I_{\perp}$.
These behaviors are illustrated on Fig. \ref{fig:scenario2}, where
the three curves show the evolution of the number of atoms on a log
scale for three different values of $I_{\perp}$. $I_{\parallel}$
and $\Delta_{\perp}$ are as in Fig. \ref{fig:fullscale}. When $I_{\perp}\lesssim I_{S}$,
the number of atoms decreases progressively until it vanishes for
$\Delta_{\parallel}\simeq-7\Gamma$. When $I_{\perp}\gg I_{S}$, the
decreasing is much faster, with a disappearance of the cloud at $\Delta_{\parallel}\simeq-5\Gamma$.
For intermediate intensities, the curve is the same as that of Fig.
\ref{fig:fullscale}: the decreasing of the number of atoms is also
fast, and the cloud disappears at $\Delta_{\parallel}\simeq-5\Gamma$.
But the curve exhibits a rebound, which means that the cloud re-appears
until it definitively disappears at $\Delta_{\parallel}\simeq-7\Gamma$.
The number of atoms in the rebound may be consequent: Fig. \ref{fig:scenario10}
shows the rebound obtained for $I_{\parallel}=9I_{S}$, $I_{\perp}=4I_{S}$
and $\Delta_{\perp}=-2\Gamma$. $N$ reaches in this case more than
1\% of the main maximum, while $\Delta_{\parallel}=-8\Gamma$. Fig.
\ref{fig:scenario10} shows that in this situation again, when $I_{\perp}$
is increased, the rebound disappears and the cloud vanishes for a
small detuning. In fact, the behavior shown on Fig. \ref{fig:scenario2}
appears to be general, whatever the values of $-2\leq\Delta_{\perp}\leq-1$
and $I_{S}\leq I_{\parallel}\leq12I_{S}$. The only explanation of
this type of behavior is that strong couplings between the different
directions arise.

\begin{figure}
\begin{centering}
\includegraphics[width=8cm]{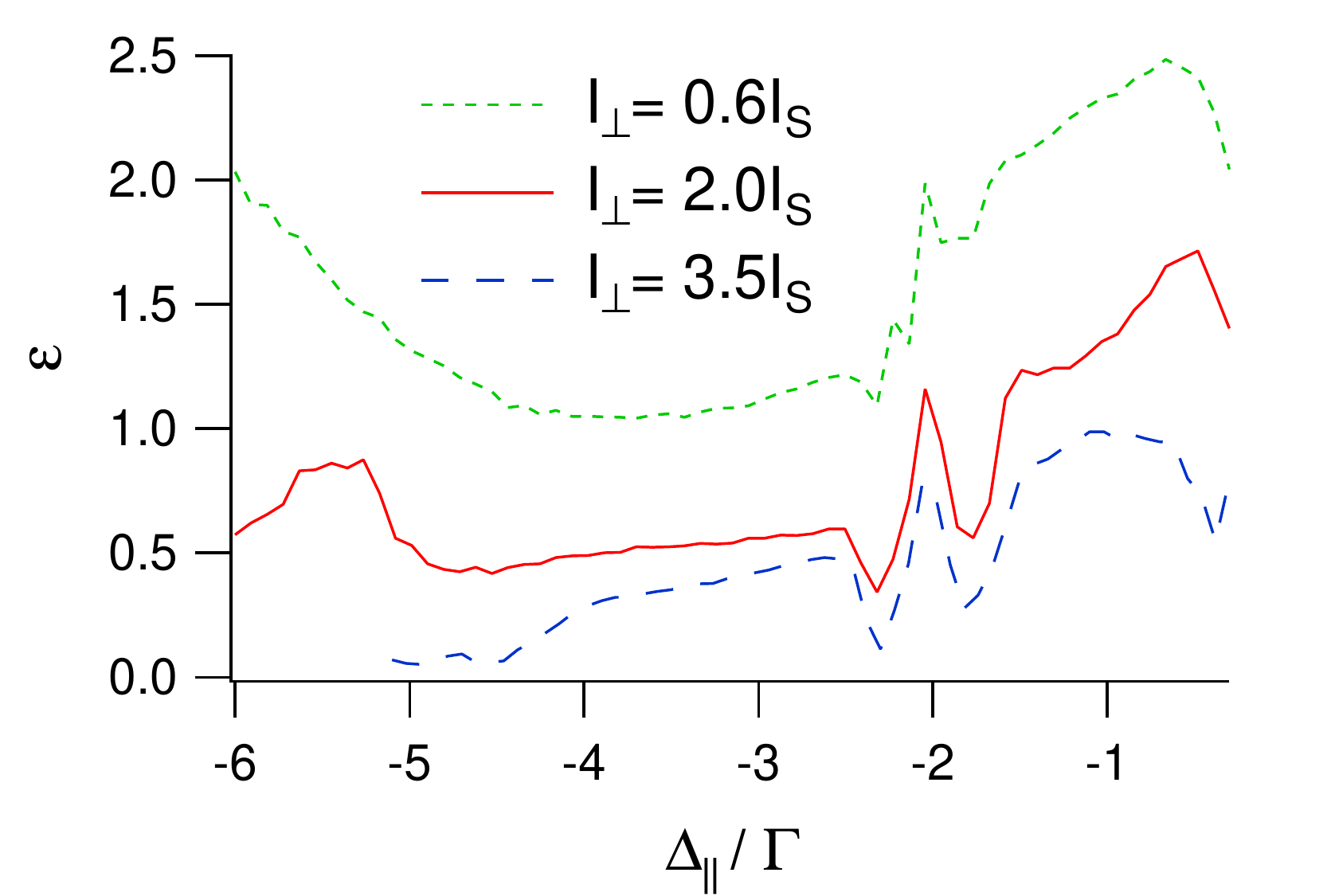}
\par\end{centering}

\caption{\label{fig:ellipticities}Ellipticity $\varepsilon$ of the atomic
cloud versus $\Delta_{\parallel}$ for $\Delta_{\perp}=-2\Gamma$,
$I_{\parallel}\simeq2I_{S}$ and different values of $I_{\perp}$.}

\end{figure}

As expected, the cloud shape depends on the trap anisotropy. Fig.
\ref{fig:ellipticities} shows the evolution of the ellipticity $\varepsilon$
for the same parameters as in Fig. \ref{fig:scenario2}. As for $N$,
the curves exhibit around the single-frequency MOT $\Delta_{\parallel}=\Delta_{\perp}$
rapid variations which probably also originate in the Raman resonance
discussed above. Except for that point, the ellipticity increases
on the blue side of the isotropic trap, before it decreases rapidly
when the resonance is closely approached. On the red side, the ellipticity
decreases slowly, and then, depending on the value of $I_{\perp}$,
it may increase again for larger detunings. We can also notice that
when $I_{\perp}$ is increased, $\varepsilon$ globally decreases,
as it could be naively expected: the larger transverse intensity compresses
the cloud in the transverse direction.

In summary of these experimental observations, several non trivial
behaviors occur in the anisotropic trap. Strong couplings between
the different directions show up. We were not able to observe instabilities
in this configuration, and we attribute it to these couplings. The
most intriguing behavior is the disappearance and re-appearance of
the cloud when $\Delta_{\parallel}$ is increased for adequate parameters.
The role of the intensities appears to be critical: the relative values
of $I_{\perp}$ and $I_{\parallel}$, as well as their values as compared
to $I_{S}$, determine the evolution of the cloud. In the next sections,
we examine how these behaviors are reproduced by different models
of the MOT.

\section{Theoretical results}

\subsection{Determination of the forces and equilibrium\label{sec:standard}}

The usual theoretical description of the MOT is based on the balance
between the different forces experienced by the atoms in the trap:
the trapping force, the shadow effect and the multiple scattering.
Expressions of these forces for an isotropic trap can be easily found
\cite{sesko1991}. 

The trapping force is the sum of the restoring force induced by the
magnetic field and the friction force induced by the light. At equilibrium,
the friction force vanishes, and the trapping force for an anisotropic
trap is: 
\begin{equation}
\mathbf{F}_{T}=-\left(\begin{array}{c}
\kappa_{x}x\\
\kappa_{y}y\\
\kappa_{z}z
\end{array}\right)
\end{equation}
where $\left(x,y,z\right)$ are the coordinates of the atom and $\kappa_{x,y,z}$
are the spring constants. Let us assume that $z$ is the coil axis,
i. e. the parallel direction. The symmetry properties of the trap
allows us to write $\kappa_{x}=\kappa_{y}=\kappa_{\perp}$ and $\kappa_{z}=2\kappa_{\parallel}$,
where the factor $2$ in $\kappa_{z}$ is arbitrary introduced so
that in the single-frequency trap, $\kappa_{\parallel}=\kappa_{\perp}$.
We have now:
\begin{equation}
\mathbf{F}_{T}=-\left(\begin{array}{c}
\kappa_{\perp}x\\
\kappa_{\perp}y\\
2\kappa_{\parallel}z
\end{array}\right)
\end{equation}

The second force is induced by the shadow effect, which results from
the intensity difference between the two counter-propagating beams,
due to the absorption. Along a given direction, this force is
\begin{equation}
F_{S}=-\frac{\sigma_{L}}{c}\left(I_{+}-I_{-}\right)\label{eq:fs}
\end{equation}
where $\sigma_{L}$ is the absorption cross section, and $I_{+}$
and $I_{-}$ the local intensities of the two beams propagating in
opposite directions. $\sigma_{L}$ depends \textit{a priori} on the
intensities and the detunings, and thus on the direction: we introduce
the spatial components $\left(\sigma_{L\perp},\sigma_{L\perp},\sigma_{L\parallel}\right)$
following $\left(x,y,z\right)$. $\mathbf{F}_{S}$ is obtained by
integration of the propagation equations for the intensities. 

The last force originates in the multiple scattering, i.e. the re-absorption
of photons already scattered by a first atom. This results in the
appearance of a repulsive force between the two atoms, written in
\cite{sesko1991} for an isotropic trap:
\begin{equation}
f_{M}=\frac{1}{4\pi r^{2}}\frac{\sigma_{R}\sigma_{L}I}{c}\label{eq:fmi}
\end{equation}
where $I$ is the total intensity, $\sigma_{R}$ the re-absorption
cross section and $r$ the distance between the two atoms. In our
case, the photon scattering by an atom leads in first approximation
to a global isotropic power equal to $2\sigma_{L\parallel}I_{\parallel}+4\sigma_{L\perp}I_{\perp}$.
Thus $\sigma_{R}$ does not depend on the direction of the scattered
photon. However, $\sigma_{R}$ depends on the frequency -- and thus
direction -- of the initial photon. Therefore, although isotropic,
different $\sigma_{R}$ are associated to the $\sigma_{L}$ of each
direction, and we obtain:
\begin{equation}
f_{M}=\frac{1}{2\pi r^{2}}\frac{\sigma_{R\parallel}\sigma_{L\parallel}I_{\parallel}+2\sigma_{R\perp}\sigma_{L\perp}I_{\perp}}{c}\label{eq:fm}
\end{equation}

The collective force $\mathbf{F}_{M}$ induced by $f_{M}$ is obtained
by integration, taking into account the atomic distribution. In \cite{sesko1991},
it is shown that for an isotropic trap, the atomic density $n$ is
constant through the whole cloud. It is easy to show that it is still
the case here: we only need to write the divergence of the three forces,
and $n$ is obtained from the equilibrium condition. We obtain a generalization
of the equation in \cite{sesko1991}:

\begin{equation}
n=\frac{c\left(\kappa_{\parallel}+\kappa_{\perp}\right)}{2\sigma_{L\perp}I_{\perp}\left(\sigma_{R\perp}-\sigma_{L\perp}\right)+\sigma_{L\parallel}I_{\parallel}\left(\sigma_{R\parallel}-\sigma_{L\parallel}\right)}\label{eq:n}
\end{equation}
$n$ being constant, the integration on the space of the propagation
equations, assuming a moderate absorption, gives:
\begin{equation}
\mathbf{F}_{S}=-\frac{2n}{c}\left(\begin{array}{c}
\sigma_{L\perp}^{2}I_{\perp}x\\
\sigma_{L\perp}^{2}I_{\perp}y\\
\sigma_{L\parallel}^{2}I_{\parallel}z
\end{array}\right)\label{eq:fsc}
\end{equation}
The integration of $f_{M}$, for any position $\mathbf{r}$ and taking
into account the ellipsoidal geometry of the cloud, is quite difficult.
However, symmetry reasons allow us to compute only the $z$ component
of the force for atoms located on the $z$ axis: 
\begin{equation}
F_{M}\left(0,0,z\right)=\frac{2n}{c}\left(2\sigma_{R\perp}\sigma_{L\perp}I_{\perp}+\sigma_{R\parallel}\sigma_{L\parallel}I_{\parallel}\right)Az\label{eq:fmc}
\end{equation}
with
\begin{eqnarray}
A & = & \left({\displaystyle \frac{\varepsilon^{2}}{\varepsilon^{2}-1}}\right)\beta\label{eq:Aexp}\\
\beta & = & \begin{cases}
{\displaystyle 1-\frac{1}{\sqrt{1-\varepsilon^{2}}}\ln\left|\frac{1+\sqrt{1-\varepsilon^{2}}}{1-\sqrt{1-\varepsilon^{2}}}\right|} & \mathrm{for}\:\varepsilon^{2}<1\\
1{\displaystyle -\frac{1}{\sqrt{\varepsilon^{2}-1}}\arcsin\left(\sqrt{\frac{\varepsilon^{2}-1}{\varepsilon^{2}}}\right)} & \mathrm{for}\:\varepsilon^{2}>1
\end{cases}\nonumber 
\end{eqnarray}

$A$ characterizes the geometry of the cloud, as it depends only on
the ellipticity. We can now write the condition of equilibrium of
the cloud on the $z$ axis, i.e. the sum of all forces equals to zero.
This condition results in a condition on $A$:
\begin{equation}
A=\frac{\kappa_{\parallel}}{\kappa_{\perp}+\kappa_{\parallel}}\left(1+\frac{\sigma_{L\parallel}^{2}I_{\parallel}\kappa_{\perp}-2\,\sigma_{L\perp}^{2}I_{\perp}\kappa_{\parallel}}{\kappa_{\parallel}\left(2\sigma_{R\perp}\,\sigma_{L\perp}I_{\perp}+\sigma_{R\parallel}\,\sigma_{L\parallel}I_{\parallel}\right)}\right)\label{eq:Atheo}
\end{equation}
If we determine the different cross sections and spring constants,
we will be able to evaluate Eq. \ref{eq:Atheo}, and to compare it
to the experimental measurements through Eq. \ref{eq:Aexp}. This
is done in the next section.

\subsection{The 1D MOT}

To determine $\sigma_{R}$, $\sigma_{L}$ and $\boldsymbol{\kappa}$,
we use the usual approximation, which considers three independent
1D MOTs perpendicular to each other. This situation has already been
studied with several levels of approximations \cite{pohl2006,romain2011},
but the studies in \cite{romain2011} are the only ones, to our knowledge,
which take into account the saturation of the transition, while the
others are always in the limit of small intensities. In \cite{romain2011}
a 1D MOT is considered: two counter-propagating laser beams with opposite
circular polarizations interact with the atoms. The atoms are the
simplest ones for which the magneto-optical trapping is possible:
the laser frequency is tuned in the vicinity of a $J=0\rightarrow J=1$
transition. In this model, the expression of $\sigma_{L}$ is:

\begin{equation}
\sigma_{L}=\sigma_{0}\frac{\Gamma^{2}}{4\Delta^{2}+2\Omega^{2}+\Gamma^{2}}\label{eq:sl1}
\end{equation}
where
\[
\sigma_{0}{\displaystyle =\frac{\hbar k_{L}\Gamma c}{2I_{S}}=\frac{3\lambda^{2}}{2\pi}}
\]
is the absorption cross section at resonance in the weak saturation
regime and $\Omega=\sqrt{\left|\Omega_{+}\right|^{2}+\left|\Omega_{-}\right|^{2}}$
is the total Rabi frequency. The individual Rabi frequencies $\Omega_{\pm}$
are defined as usual:
\[
\frac{\Omega_{\pm}^{2}}{\Gamma^{2}}=\frac{I_{\pm}}{2I_{S}}
\]

Different expressions of $\sigma_{R}$ have been obtained in \cite{romain2011},
depending on the relative values of $\Delta$, $\Omega$ and $\Gamma$.
For example, in the very common experimental situation where $\left|\Delta\right|\gg\Omega\gg\Gamma$,
its expression is:
\begin{equation}
\sigma_{R}=\frac{\sigma_{0}}{8}\frac{\Omega^{2}}{\Delta^{2}}\label{eq:sr}
\end{equation}

The last quantity to evaluate is the spring constant. The expression
of $\mathbf{F}_{T}$ obtained in \cite{romain2011} allows us to find
an expression similar to that of the friction in \cite{dalibard1984}:
\begin{eqnarray}
\kappa & = & -8\mu_{B}bk_{L}\frac{\Gamma\Delta}{\left(4\Delta^{2}+2\Omega^{2}+\Gamma^{2}\right)^{2}}\nonumber \\
 &  & \times\left[\Omega^{2}+\frac{16\left|\Omega_{+}\right|^{2}\left|\Omega_{-}\right|^{2}\Gamma^{2}}{16\Gamma^{2}\Delta^{2}+\left(2\Gamma^{2}+\Omega^{2}\right)^{2}}\,\left(1-\frac{\Omega^{2}}{4\Gamma^{2}}\right)\right]\label{eq:kappa}
\end{eqnarray}

\begin{figure}
\centering{}\includegraphics[width=8cm]{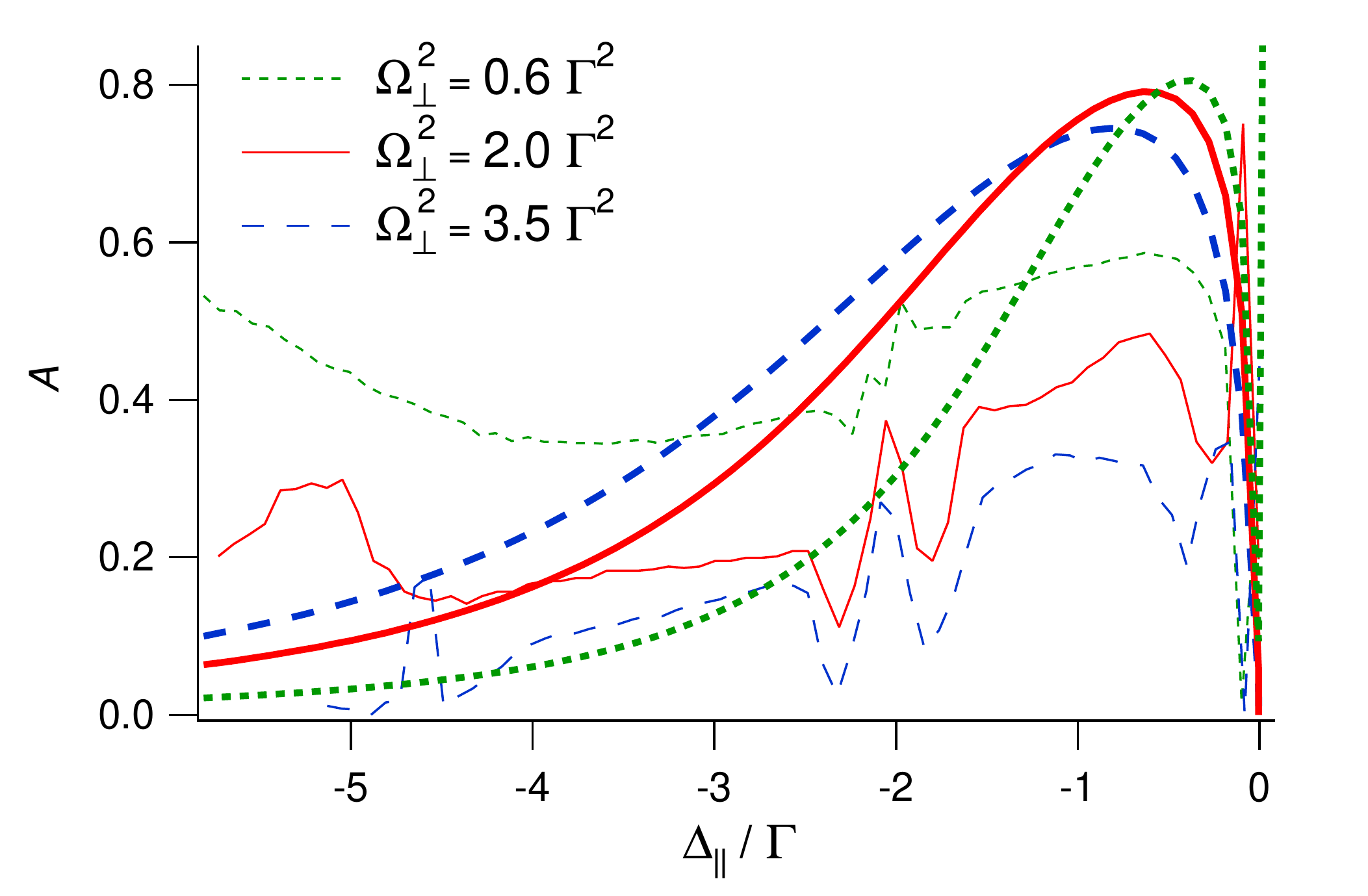}\caption{\label{fig:comparaison}$A$ parameter versus $\Delta_{\parallel}$
for $\Delta_{\perp}=-2\Gamma$, $I_{\parallel}\simeq2I_{S}$ and different
values of $I_{\perp}$. Parameters are the same as in Fig. \ref{fig:ellipticities}.
Thick curves correspond to theoretical values, while thin curves are
the experimental records.}
\end{figure}
We are now able to calculate $A$ both experimentally with the Eq.
\ref{eq:Aexp}, and theoretically with Eq. \ref{eq:Atheo}. Fig. \ref{fig:comparaison}
shows as an example the curves obtained for the same parameters as
in Fig. \ref{fig:scenario2} and \ref{fig:ellipticities}. There is
no agreement between the experimental curves and the theoretical ones.
The main discrepancy concerns the evolution of the $A$ parameter
as $I_{\perp}$ is changed: experimentally, $A$ decreases as $I_{\perp}$
is increased, while it is the contrary in the model. Another major
inconsistency lies in the fact that for large detunings, the theoretical
evolution of all quantities, i.e. $A$, $\kappa$ (Fig. \ref{fig:kappa},
black bold solid line), $\sigma_{L}$ and $\sigma_{R}$, is monotonic.
Such a monotonic evolution cannot explain the disappearance of the
atomic cloud for intermediate values of $\Delta_{\parallel}$, as
observed in the experiments, neither the re-appearance for larger
values.

Thus it is clear that the present model is not able to reproduce the
experimental observations. Let us remember that in this model, several
known physical phenomena have been neglected. The question is now
which of them plays finally a role which has been underestimated.
The experimental observations shows that the relative values of $I_{\perp}$
and $I_{\parallel}$ play a crucial role in the evolution of both
the ellipticity and the number of atoms. As in the present model,
the 3D MOT is approximated to three perpendicular independent 1D MOTs,
such effects can not be found. Thus it seems logical to enhance the
present model to take into account the cross intensity effects between
the parallel and transverse beams, and in particular the cross saturation
effects. In the next section, we modify the standard model to take
into account these effects.

\subsection{Introduction of couplings between MOT arms}

In the above model, the expression of the parameters $\sigma_{R}$,
$\sigma_{L}$ and $\boldsymbol{\kappa}$ result from a 1D approximation.
We would like here to enhance this model to take into account, at
least partly, the effects induced by the couplings of each pair of
beams with its transverse ones. Building a real 3D model would be
rather complex. An intermediate model consists in still considering
three 1D MOTs, but to introduce for each beam a correction induced
by the two other pairs of beams. However, considering the effects
of the transverse beams implies in our case to study the excitation
of the atomic transition by two quasi-resonant fields with similar
amplitudes. The theoretical description of this problem is still laborious.
However, it can be greatly simplified for the parallel direction.
Indeed, in this case, the two transverse beams have the same frequency,
and we can still simplify the model if we consider that the four transverse
beams are linearly polarized along $z$ rather than circularly polarized.
In this case, the $\sigma^{+}-\sigma^{-}$ longitudinal beams interact
with the $\left|m=\pm1\right\rangle $ levels, while the transverse
beams, $\pi$ polarized, interact only with the $\left|m=0\right\rangle $
level. In this case, the calculations can be done in the same way
as in \cite{romain2011}. The resulting expressions are rather complex
and their expressions are useless to understand the underlying physics.
However, we expect that, as the fundamental level is now coupled to
the $\left|m=0\right\rangle $ level, the number of atoms susceptible
to absorb the $\sigma^{+}-\sigma^{-}$ photons is smaller. This should
lead to a decreasing of the absorption cross section, and it is effectively
what we observed.

\begin{figure}
\centering{}\includegraphics[width=8cm]{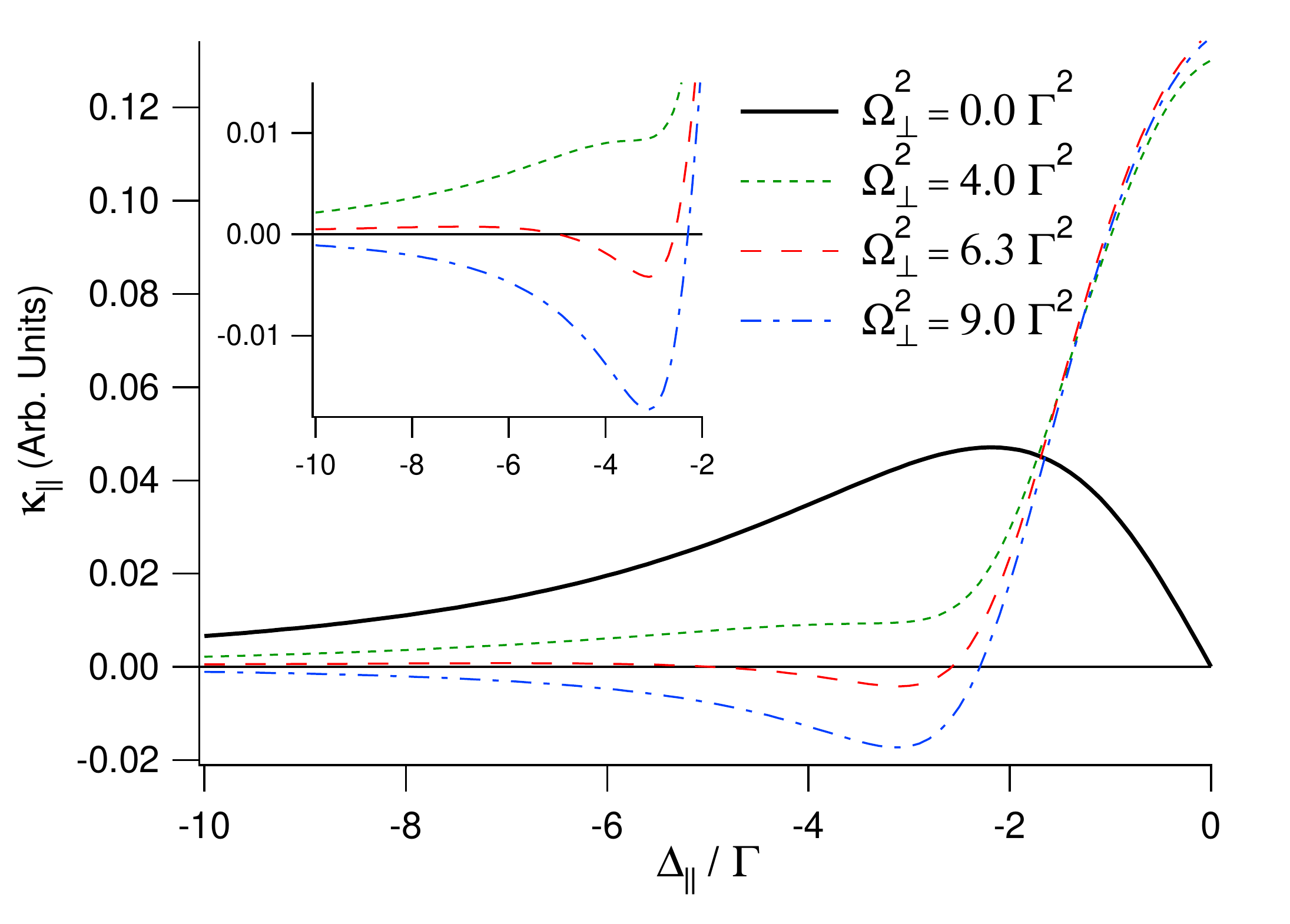}\caption{\label{fig:kappa}Plot of $\kappa_{\parallel}$ versus $\Delta_{\parallel}$
for different values of $\Omega_{\perp}^{2}$. Parameters are $\Delta_{\perp}=-2\Gamma$
and $\Omega_{\parallel}^{2}=9\Gamma^{2}$.}
\end{figure}
How the evolution of the spring constant is changed when the transverse
coupling is taken into account is more difficult to predict intuitively.
And indeed, this evolution is more complex, as shown on Fig. \ref{fig:kappa},
where the values of $\kappa_{\parallel}$ are plotted versus $\Delta_{\parallel}$
for different values of $I_{\perp}$. When $I_{\perp}$ is taken into
account, the shape of the curve changes radically. First, because
of the coupling, the spring constant is no more zero when the parallel
beam is at resonance. For small $I_{\perp}$ (Fig. \ref{fig:kappa},
green dotted line), $\kappa_{\parallel}$ is always positive, and
the atoms are trapped whatever $\Delta_{\parallel}$. Thus when the
detuning is increased, the cloud population decreases monotonically
until it vanishes for large detuning. For larger $I_{\perp}$ (red
dashed line), $\kappa_{\parallel}$ becomes negative for intermediate
values of $\Delta_{\parallel}$, and becomes positive again for large
detuning. In the interval where $\kappa_{\parallel}$ is negative,
the atoms are repelled from the trap, and thus for these intermediate
detunings, the atomic cloud disappears, but re-appears at large detunings.
Finally, for even larger $I_{\perp}$ (blue dotted dashed line), $\kappa_{\parallel}$
becomes negative for intermediate values of $\Delta_{\parallel}$,
and remains negative for large detunings: the cloud disappears for
rather small $\Delta_{\parallel}$. This behavior is fully consistent
with the experimental observations, and thus we can already say that
the introduction of cross saturation effects allows us to explain
the behavior of an anisotropic trap.

Note also on Fig. \ref{fig:kappa}, the value of the spring constant
is also changed for the balanced single-frequency trap: it is for
example decreased by a factor 2.6 for $\Delta_{\parallel}=-2\Gamma$
and $\Omega_{\parallel}^{2}=\Omega_{\perp}^{2}=9\Gamma^{2}$. This
shows that although usually neglected, the transverse couplings between
the trap arms could change significantly the trap parameters.

\begin{figure}
\centering{}\includegraphics[width=8cm]{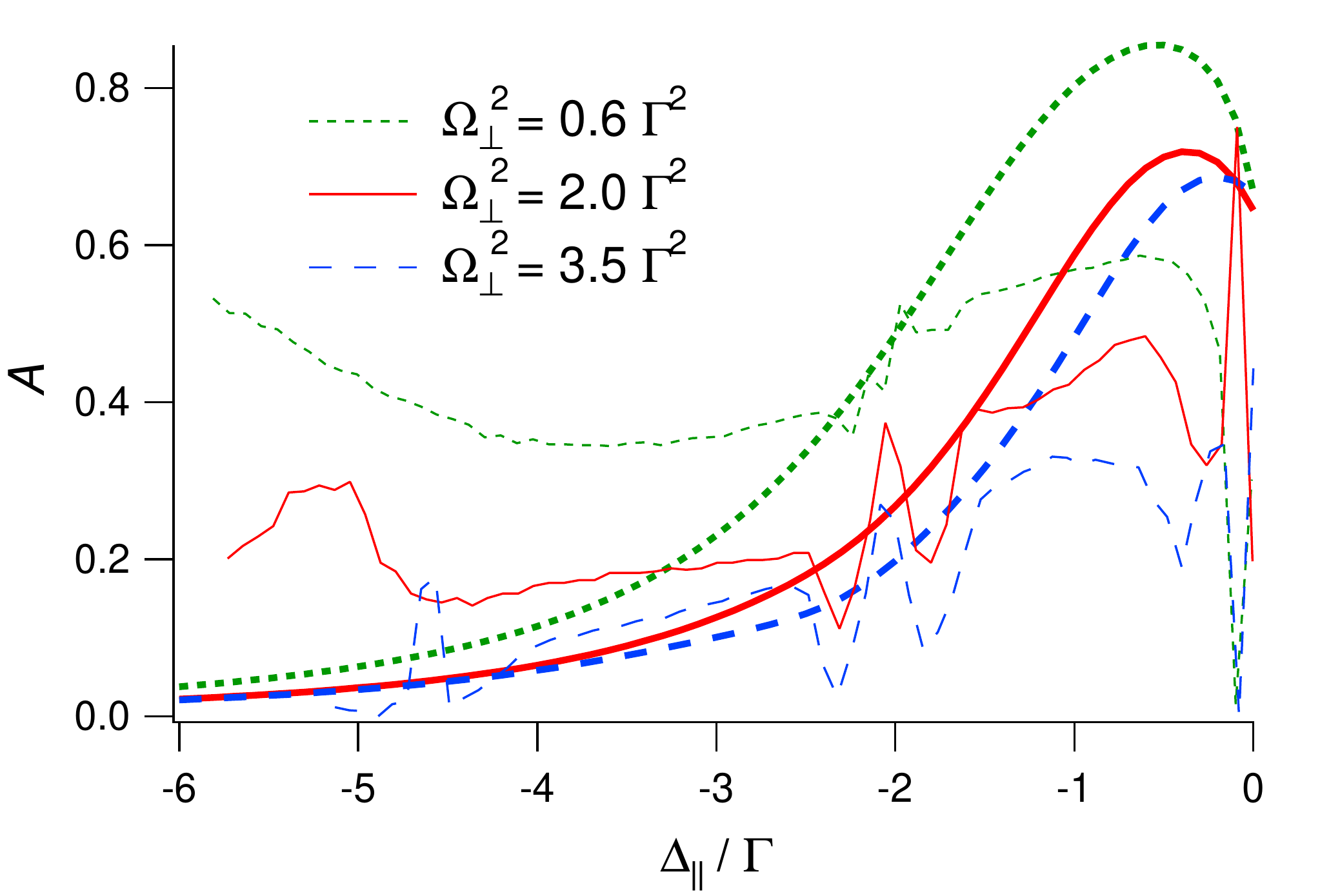}\caption{\label{fig:comparaisonsc}$A$ parameter versus $\Delta_{\parallel}$
for $\Delta_{\perp}=-2\Gamma$, $I_{\parallel}\simeq2I_{S}$ and different
values of $I_{\perp}$. Parameters are the same as in Fig. \ref{fig:comparaison}.}
\end{figure}
We also noticed in section \ref{sec:standard} that the standard model
was unable to reproduce the global evolution of the ellipticities
when the transverse intensities are changed. To check that point,
we plotted the same curves as in Fig. \ref{fig:comparaison}, but
for the cross saturation model (Fig. \ref{fig:comparaisonsc}). The
global evolution of the ellipticities, both theoretical and experimental,
are now qualitatively consistent. In particular, when $I_{\perp}$
increases, $A$ decreases in both cases.

\section{Conclusion}

In this paper, we study the behavior of a dual frequency anisotropic
MOT. Experimental measurements show several counter-intuitive behaviors,
in particular an interval of detunings where atoms are no more trapped,
while they are trapped for smaller and larger detunings. We show that
the usual model, which neglects the cross saturation effects, is unable
to explain this behavior. We build a model taking into account these
cross saturation effects, and show that this model leads to behaviors
similar to the experimental ones. The agreement between the experiments
and the model is only qualitative. It is not surprising, as numerous
approximations remain in the new model. However, it shows that cross
saturation effects play a key role in this system.

We also show that even in the traditional balanced single-frequency
trap, the couplings between the different arms of the trap change
significantly the trap characteristics. As a consequence, these effects
should be taken into account when a detail understanding is required. 

The coupling between the different directions of the trap also makes
it difficult to separate the dynamics of the MOT along the different
directions. Thus confining the instabilities of the 3D MOT in only
one direction appears to be unrealistic, and a full 3D dynamical model
of the MOT seems now necessary.

\end{document}